# Synthesis and physical properties of LiBC intermetallics

Dmitri Souptel [a) *], Zakir Hossain [b)], Günter Behr [a)], Wolfgang Löser [a)], Christoph Geibel [b)]

[a)] *Leibniz-Institute for Solid State and Materials Research Dresden, PF 27 01 16, D-01171 Dresden, Germany*

[b)] *Max-Planck-Institute for Chemical Physics of Solids, Nöthnitzer Str. 40, D-01187 Dresden, Germany*

**Abstract:**

Polycrystalline samples of LiBC compounds, which were predicted as possible candidate for high-$T_c$ superconductivity, have been synthesised by a flux method and investigated by means of electrical resistivity and magnetic susceptibility. Scanning electron microscopy and X-ray diffraction patterns showed a plate-like morphology and a single-phase nature of LiBC samples for starting composition of $Li_{1.25}BC$ (flux composition). The lattice constants *a*, *c* display a systematic variation with *x* and has maximum volume of the hexagonal unit cell at $x = 1.25$. Electrical resistivity measurements revealed an extrinsic semi-conducting behaviour of the single-phase LiBC with an activation energy of 18 meV and a maximum specific resistivity of 2.5 Ωcm at 300 K. In contrast to the theoretical prediction of high $T_C$, no superconducting features were detected down to 2 K both, by measurements of electrical resistivity and magnetic susceptibility.

PACS: 72.20.-i, 81.05.Zx, 81.20.Ka

*Keywords:* A. Semiconductors, A. Superconductors, B. Crystal growth, D. Electronic transport

## 1. Introduction:

In the past two years there were many efforts to understand the unexpected superconducting behaviour of $MgB_2$ [1, 2, 3], which led to a growing interest to closely related systems especially to the layered compounds with metallic behaviour or compounds which become metallic by doping [4, 5, 6, 7].

LiBC is an insulating material [6, 8] with a gap that arises from the replacement of $B_2$ by BC. But, there are reports [5, 6] about possibilities of heavy hole doping of this compound by varying the Li concentration but retaining its crystal structure. The calculations of electronic structure of isoelectronic compounds [4, 5, 7] gave rise to the conclusion that LiBC is a potential superconducting material. Moreover, the calculated $T_C$ of $Li_xBC$ for $x$ = 0.5 is about 100 K, and for x = 0.25 at least 40 K [5]. For the first time this compound was prepared by Wörle et.al. [6] who also determined some of their physical and chemical properties. LiBC crystallised in primitive hexagonal lattice with space group $P6_3/mmc$ and lattice parameters a = 0.2752 and c = 0.7058 nm. The B and C atoms form a planar so-called heterographite layer [6, 8, 9]. The interlayer regions are filled by Li. This crystallographic structure is similar to that of $MgB_2$ and the interlayer spacing is nearly identical. The BC layers are stacked in an alternating fashion. Pronin et al. investigated some electronic and optical properties of LiBC single crystals and revealed a B-C stretching mode expected to drive phonon-mediated superconductivity [10].

The aim of this report is a synthesis of Li deficient LiBC compounds from melt fluxes $Li_xBC$ with various Li portions $x$ and different process parameters, and to provide a survey of their low-temperature magnetic and transport properties in a temperature interval from 2 to 300 K.

## 2. Experimental

The Li granules (99.4%, MaTeck), crystalline (99.52%, Eagle Picher) or amorphous (97%, Heraeus) powder of B and graphite powder (99.9% MaTeck) were used as starting materials for synthesis of $Li_xBC$ samples. The elements were weighted in desired ratio, mixed, pressed into pellets 9 mm in diameter and placed in Ta crucible 10 mm in diameter and 60 mm length with 0.5 mm thick walls. Then crucible was closed with the Ta cap and sealed in arc melting furnace. All processes were carried out in a protecting Ar gas flow in order to prevent oxidization of the materials. Then the crucibles were placed into tube furnace (Nabertherm) and heated in Ar atmosphere. The temperature course of the heat treatment (homogenization 1 h at 500°C, heating up with 200 K/h and holding 1 h at 1500°C, and fast cooling with 550 K/h to room temperature) was almost the same as that described in Ref. [6] . After this procedure the crucibles were opened in air and inserted into quartz ampoules of about 20 cm length evacuated up to $10^{-3}$ mbar and sealed. For distillation of the Li excess those parts of the ampoules containing the Li-B-C samples were heated to about 850°C for 4 h. The other end of the ampoules were maintained at room temperature where the evaporated Li was precipitated at the cool quartz walls.

A series of starting flux compositions $Li_xBC$ was prepared for synthesis of LiBC samples with $x$ = 0.5; 0.75; 1.0; 1.25; 2.0 and 2.4 in order to elucidate the possibility of the LiBC synthesis with different Li deficiencies. All the samples obtained were porous, containing LiBC particles with color from golden to brown and black depending on the starting composition. Several samples were annealed at 900°C, 1000°C and 1100°C for 20 h in closed evacuated quartz ampoules to test the thermal stability of the LiBC and eventually inducing Li-deficiency of the LiBC compound.

The constituent phases of the samples and lattice parameters were revealed by means of powder x-ray diffraction analysis with STOE automated multipurpose powder diffractometer (STADIP-MP). The resistivity measurements were carried out using AC transport option of Physical Property Measurement System (PPMS, Quantum Design) by a standard low frequency AC four probe technique in a temperature range 1.8 to 300 K. Magnetic measurements have been performed using a commercial Superconducting Quantum Interference Device (SQUID) magnetometer in the temperature range 2 – 300 K.

## 3. Results and discussion

### 3.1 XRD measurements

In Fig. 1 the SEM images of samples with x = 1.0; 1.25; 2.0 are shown. The best samples containing platelets of 2 µm thickness and 20 µm in diameter with golden colour and clean surfaces were obtained for a starting flux composition $Li_xBC$ with x = 1.25 (Fig. 1a). Whereas, for x < 1.25 the plate-like morphology was changed into elongated fine grains (Fig. 1b) of black color, which may represent Li-depleted particles according to Wörle [6]. For flux compositions x > 1.25 some (carbon) fibres or tubes have been detected between the platelets (Fig. 1c). Heat treatment of samples at 1000°C led to a partial decomposition and to small particles on the LiBC platelets (Fig. 1d). The corresponding XRD patterns in Fig. 2 confirmed the single phase nature of samples prepared for flux compositions with x = 1.25. All the other samples contain impurity phases to some extend, which have not been identified completely. For x = 2.0 strong reflections of graphite were detected, which confirm the suggestion of carbon fibres from the SEM image. Weak reflections of boron rich phases $BC_x$ were also detected in samples prepared from Li-depleted fluxes. The obtained LiBC samples crystallized in primitive hexagonal $P6_3/mmc$ structure. The lattice parameters a, c and the unit cell volume are derived from X-ray diffraction patterns. A characteristic dependence on the flux composition $Li_xBC$ is shown in Fig. 3. The trend is increasing of the unit cell volume with increasing of Li content in $Li_xBC$ flux composition until a maximum at $x = 1.25$ is achieved. The change of lattice constant is more apparent for *a* than for the distance of the layers *c*. We may attribute the observed change of the lattice to a defect structure and a finite homogeneity region of the LiBC compound. The lattice constants at $x = 1.25$, a = 0.275 nm and c = 0.7055 nm are in fair agreement with a = 0.2752 nm and c = 0.7058 nm reported by Wöhrle at al. [6] for the stoichiometric LiBC compound. Whereas, our minimum values achieved for $x = 2.4$ a = 0.2750 nm and c = 0.7042 nm are still far from a = 0.2731 nm and c = 0.7006 nm reported for Li-depleted brown crystals in Ref. [6]. Annealing of off-stoichiometric samples (from flux compositions $x = 2.0$ and 2.4) at 900°C increased the unit cell volume by a sizeable amount. That may hint to a relaxation of the defect structure. The effect of annealing was not further enhanced by a temperature rise to 1000°C. In accordance with [6] we suppose that a Li-depletion of the LiBC compound is responsible for the shift in lattice constants and the reduction of the unit cell volume. However, because of difficulties of concentration determination of the composing light elements it is not possible to confirm this supposition by other methods such as electron microprobe measurements. Attempts to reduce Li contents of crystals by annealing and maintaining the same crystal structure were not successful. LiBC decomposes by formation of primitive borides and for > 1000°C it decays into the elements. The higher the annealing temperature the weaker are X-ray reflections of the LiBC compound in the annealed samples. For 1100°C almost no LiBC reflections were observed. Therefore we must suppose that the annealing is not an appropriate technique for reducing Li contents in the LiBC compound.

In this study we have focused our attention to measure the physical properties of Li-deficient samples prepared from different flux compositions $Li_xBC$ with $x = 0.5$ to 2.4. As stated above only for $x = 1.25$ single phase LiBC samples were obtained, whereas for all other samples we are faced with impurity phases to some extent as shown in the X-ray diffraction pictures Fig. 2.

### 3.2 Resistivity measurements

The electrical resistivity of all the distilled LiBC samples prepared from $Li_xBC$ fluxes of different compositions from 0.5 to 2.4 were measured in the range 1.8 - 300 K. Results for x = 1.0, 1.25 and 2.4 are presented in Fig. 4. The single-phase sample x = 1.25 exhibits semiconducting behaviour and an outstanding high resistivity with 2.5 Ωcm near room temperature. The semiconducting behaviour is less pronounced and the magnitude of resistivity is reduced by 2 - 3 orders of magnitude for samples prepared from fluxes with $x \neq 1.25$ as well as in annealed samples. In any case no superconducting transition to zero resistance was observed throughout the temperature range from 1.8 K to 300 K. The semiconducting behaviour of the single-phase sample ($x = 1.25$) resembles the overall behaviour reported in ref. [10]. However, different from those authors we have analysed the data in the common *Log r vs. (1/T)* plot (inset to Fig. 4). The small activation energy of *DE* = 18 meV derived from the linear extrapolation at temperatures 140 < T < 300 K is in accordance with an extrinsic semiconducting behaviour. The electrical behaviour of single-phase LiBC samples strongly resembles that of doped and compensated semiconductors [10] where the activation energy near room temperature determined by the difference $E_F$ - $E_D$ between Fermi energy $E_F$ and the level of dopant $E_D$. For low temperatures T < 20 K there is an obvious transition toward another conductivity mode. Therefore, the low temperature specific resistivity was fitted $\rho \sim \exp(B/T^{1/4})$ valid for variable range hopping [11]. As shown in Fig. 5 the experimental results of the ρ vs. $T^{1/4}$ plot reasonably well match with a linear behaviour. The linear slope is decreased in the sequence of B = 0.31 $K^{1/4}$ for single-phase samples (x = 1.25) to B = 0.19 $K^{1/4}$ (x = 2.4) and B = 0.12 $K^{1/4}$ (x = 1.0). The same sequence also holds for the reduction of the magnitude of specific resistivity of samples grown from fluxes with x = 1.0 and 2.4 shown in Fig. 4. Very small activation energies *DE* ≤ 0.1 meV at room temperature were derived for those samples. Therefore, we suppose that they approach to a nearly metallic behaviour because of an enhanced density of defects. Wörle et al. [6] have attributed the variation of room-temperature specific conductivity of LiBC samples by several orders of magnitude to both doping by the stoichiometry deviations of the LiBC intermetallic compound and grain boundary effects. Since some samples (with $x \neq 1.25$) contain graphite as a minority phase, the latter can contribute to the metallic behaviour, too. However, the minority phase does not form a continuous network according to our SEM images (Fig. 1c,d) and will not dominate the electric properties. Therefore, enhanced doping is presumed to be responsible for the enhanced electric conductivity in samples with $x \neq 1.25$.

### 3.3 Magnetization measurements

The magnetic measurements in all cases exhibit a small positive magnetization (M > 0) characteristic for paramagnetic behaviour. However, we have not found any diamagnetism associated with superconductivity for all LiBC samples investigated down to temperatures of 2 K. In Fig. 6 the magnetization of the single phase LiBC sample (x = 1.25) between 300 and 2 K is shown under zero-field cooled and field cooled conditions with magnetic field 2 mT. Both data sets are recorded during warming up of the samples. There is a sizeable difference between susceptibility for the zero-field cooled and field cooled conditions below a bifurcation temperature $T_b \approx 250$ K. It results from thermally activated magnetization processes due to the presence of lattice sites carrying magnetic moments. In field cooled samples the frozen-in magnetic order is reduced and magnetization continually slows down with rising temperature. For zero-field cooled samples, on the other hand, with increasing temperature a (weak) magnetic order of the interacting magnetic moments is established by thermally activated processes until a maximum is achieved near 50 K. With further increase of temperature the magnetization is again reduced by thermal fluctuations. The origin of the weak magnetism is not fully understood. It may be due to small magnetic impurities in nonmagnetic LiBC rather than to the LiBC compound itself.

For samples with x = 1.25 and 1.0 a sharp drop of the susceptibility is observed at ~ 5 K. This may be of superconducting origin of a minority phase in the semiconducting matrix. Since for x = 1.25 only LiBC structure was detected we could propose that this is a small amount Li deficiency phase in stoichiometric LiBC matrix.

**4. Conclusions**

The LiBC compounds were synthesised from $Li_xBC$ fluxes for a wide range of x = 0.5 to 2.4 with particular emphasis to the formation of a Li-deficient phase, which is supposed to show superconductivity. Lattice parameters and unit cell volume depend on the Li content of the flux and indeed suggest a homogeneity region of the LiBC compound. Single-phase LiBC samples were only achieved for x = 1.25. They exhibit extrinsic semiconducting behaviour over a wide temperature interval of 2 to 300 K similar to doped and compensated semiconductors. Samples prepared from other flux compositions (x ≠ 1.25) display a smaller electrical resistivity and a tendency towards metallic behaviour which is attributed to the enhanced doping level. Attempts to prepare Li-deficient samples with superconducting properties by vacuum annealing failed. Small positive magnetization was observed in the samples and ascribed to paramagnetic impurities. Both, resistivity and magnetization measurements do not hint any superconducting behaviour above 2 K. A $Li_yBC$ compound with *y* far from unity and superconducting properties predicted by theory was not achieved and possibly requires additional investigations. But, in spite of the problems with producing of LiBC doping, this compound seems to be a good candidate for field-effect doping (FED) to achieve superconductivity [12].

**Acknowledgements**

The authors would like to thank Julia Ferstl for technical support, Dmitri Jerebtsov and Arnulf Möbius for useful comments and discussions. The work was supported by the Saxon Ministry for Science and Culture (SMWK).

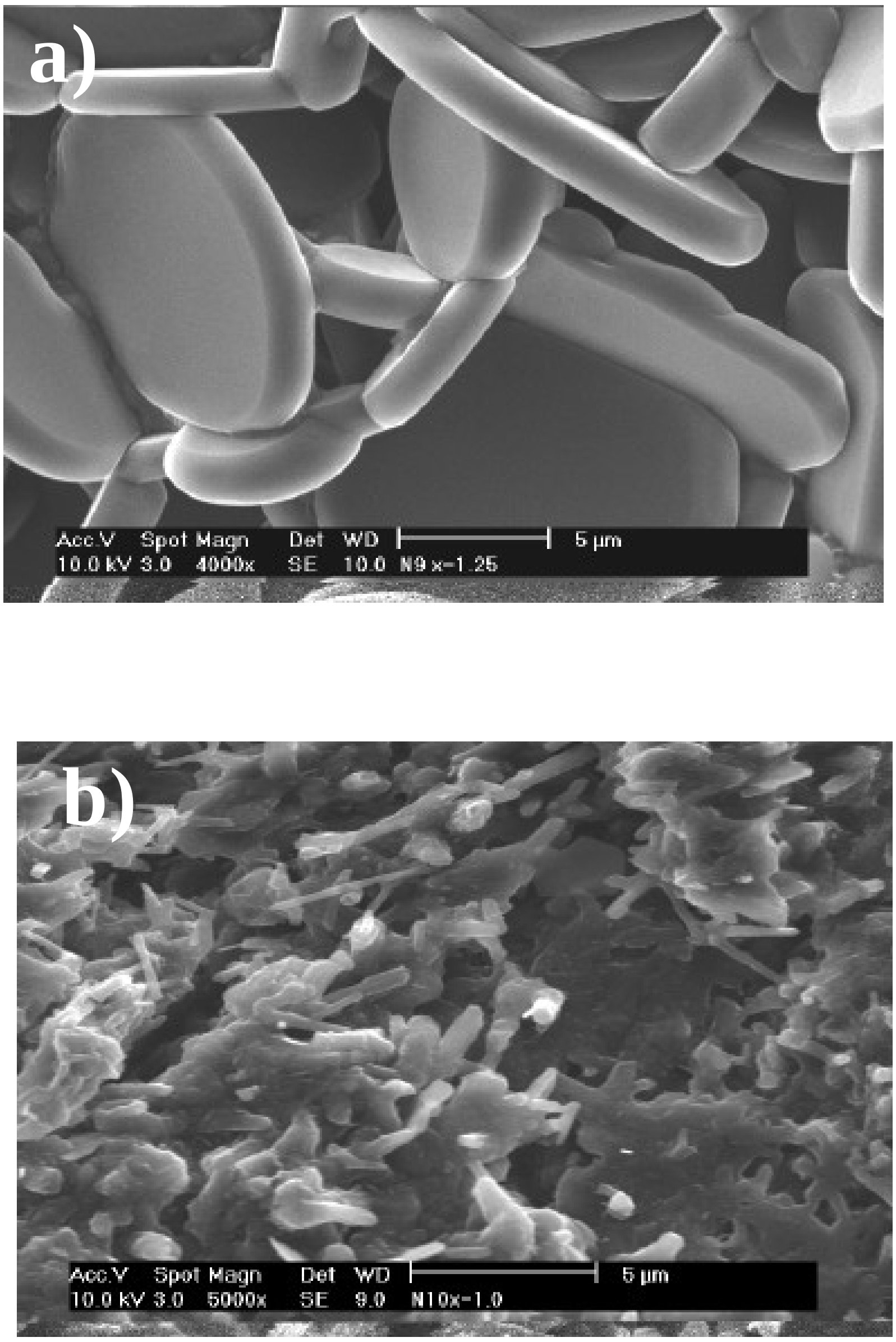

Fig. 1a, b. SEM images of LiBC samples prepared from various flux compositions $Li_xBC$.

(a) x = 1.25 and (b) x = 1.0;

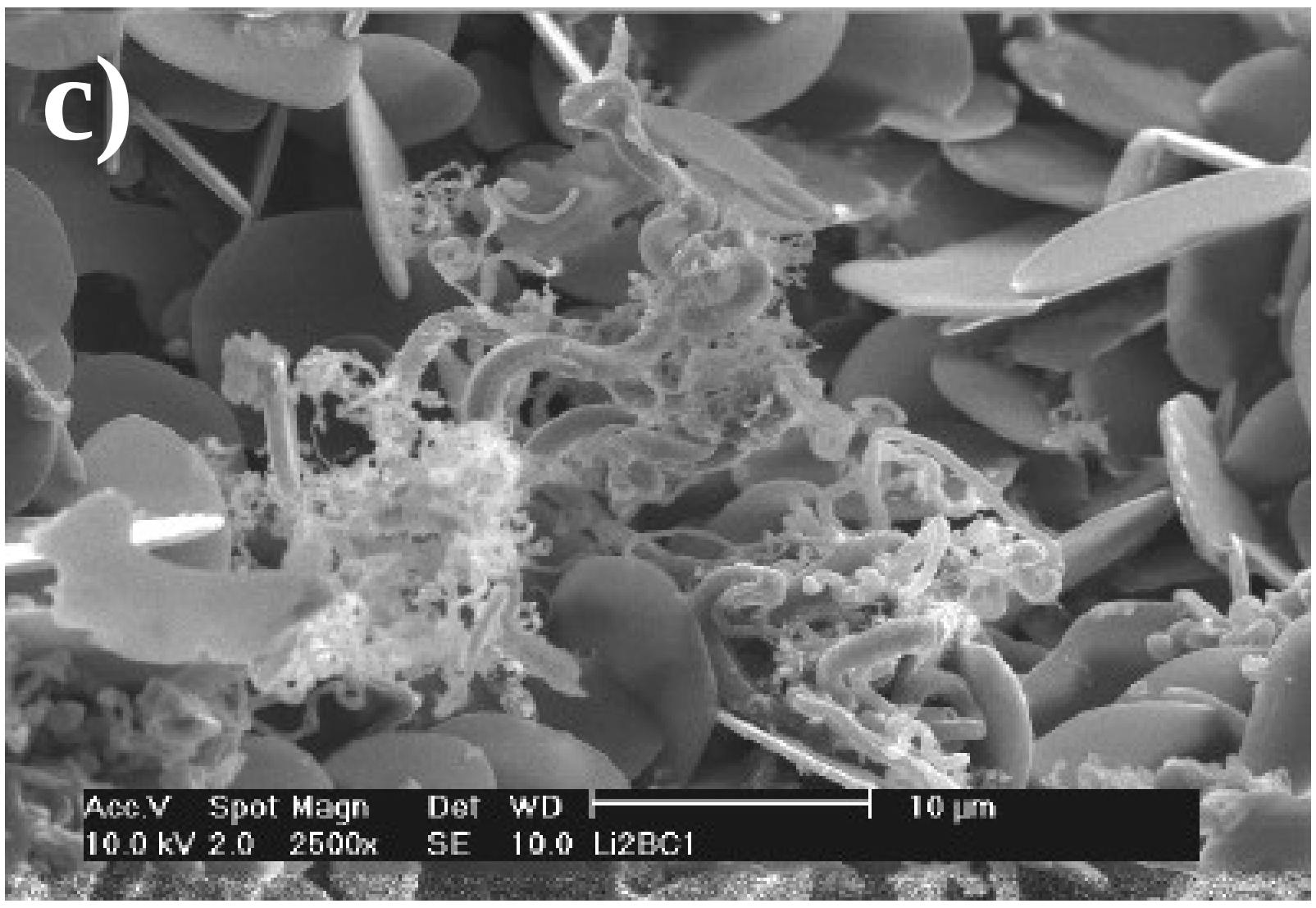

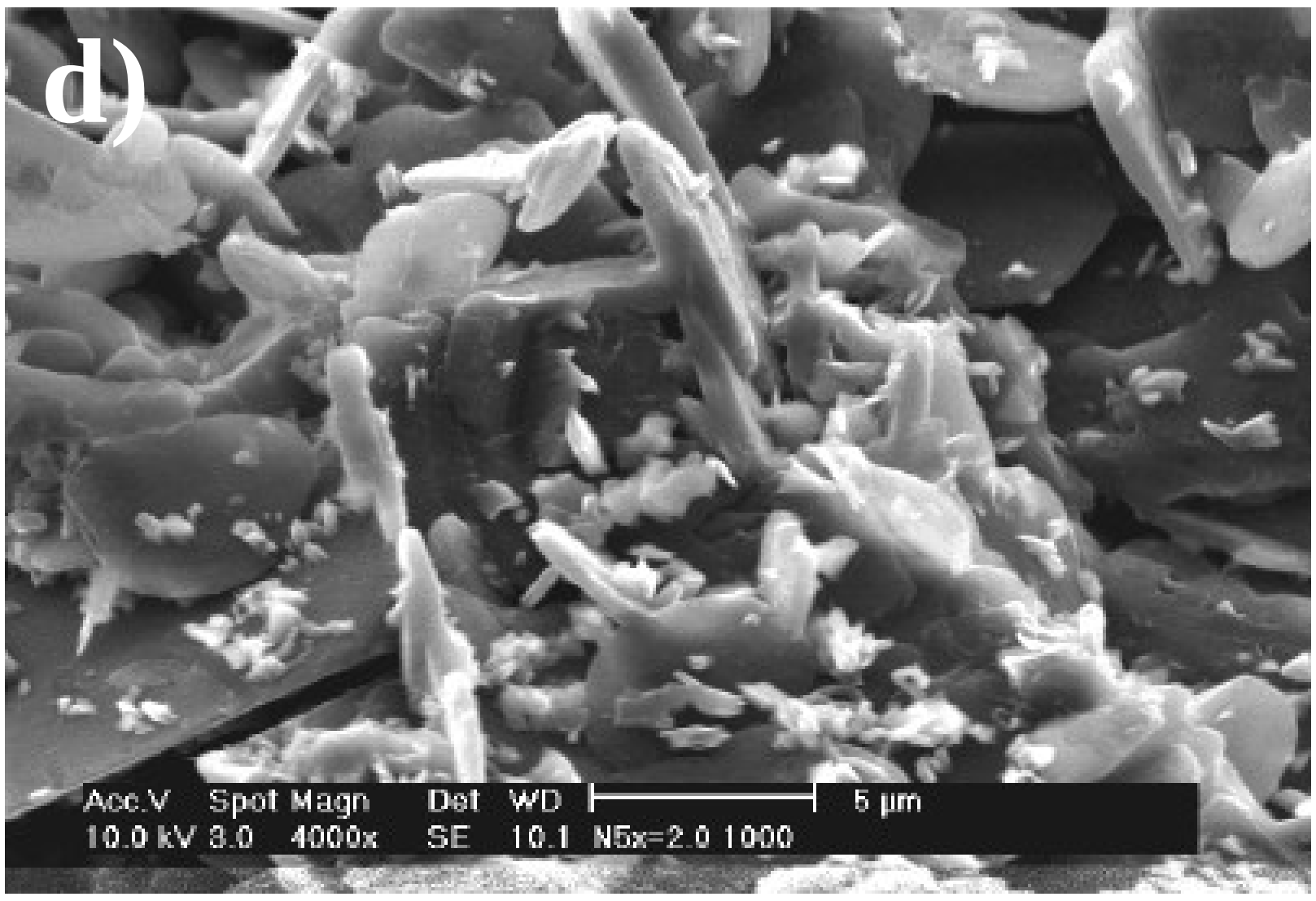

Fig. 1c, d. SEM images of LiBC samples prepared from various flux compositions $Li_xBC$. (c) x = 2.0 , tube-like inclusions and LiBC platelets; (d) x = 2.0 , after heat treatment 1000°C for 20 h.

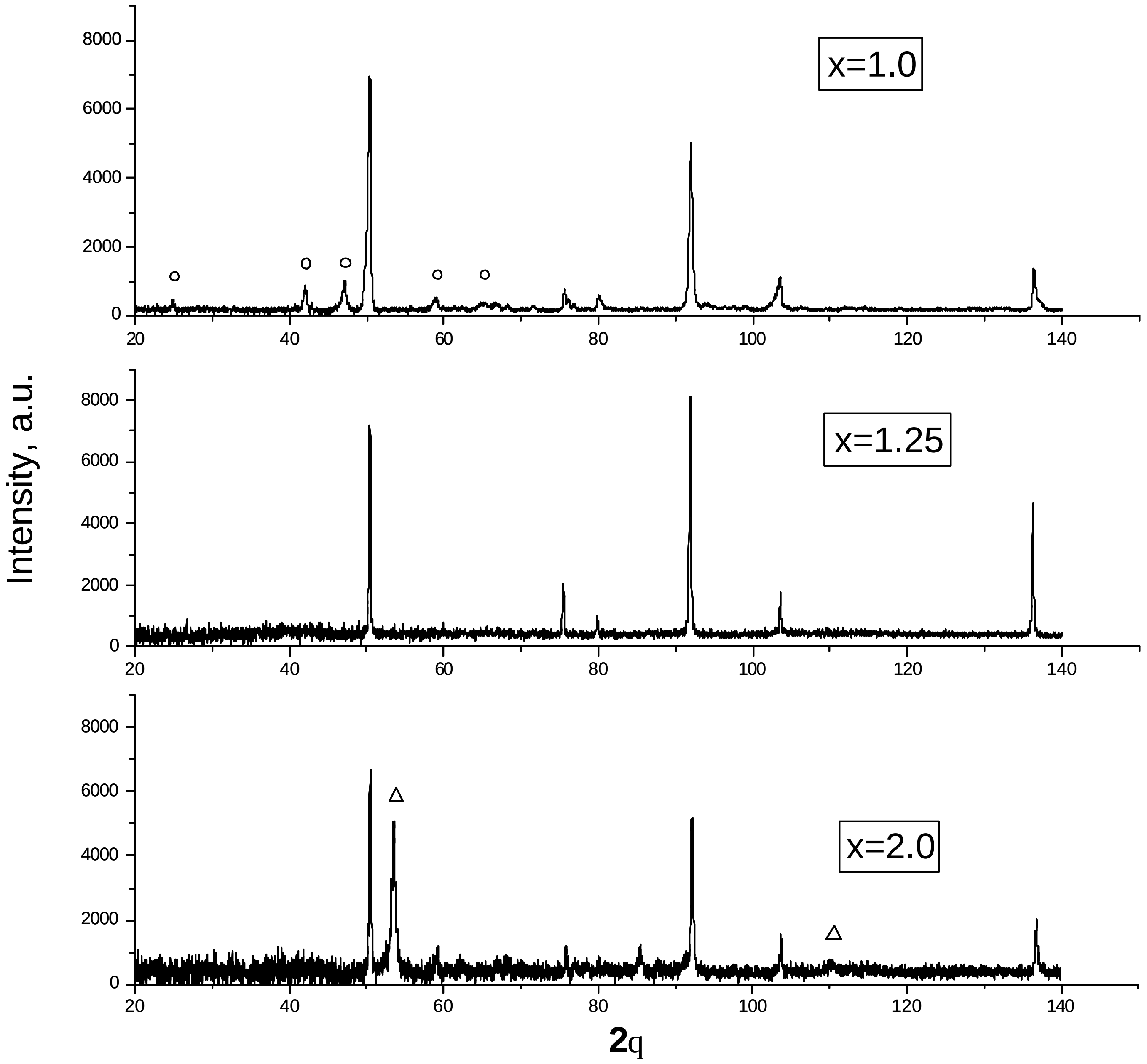

Fig. 2. XRD pattern of LiBC samples prepared from various flux compositions $Li_xBC$ for x = 1.0, 1.25 and 2.0. The reflections of the different impurity phases are indicated by full dots (boron-rich (B,C)-phase) and triangles (graphite).

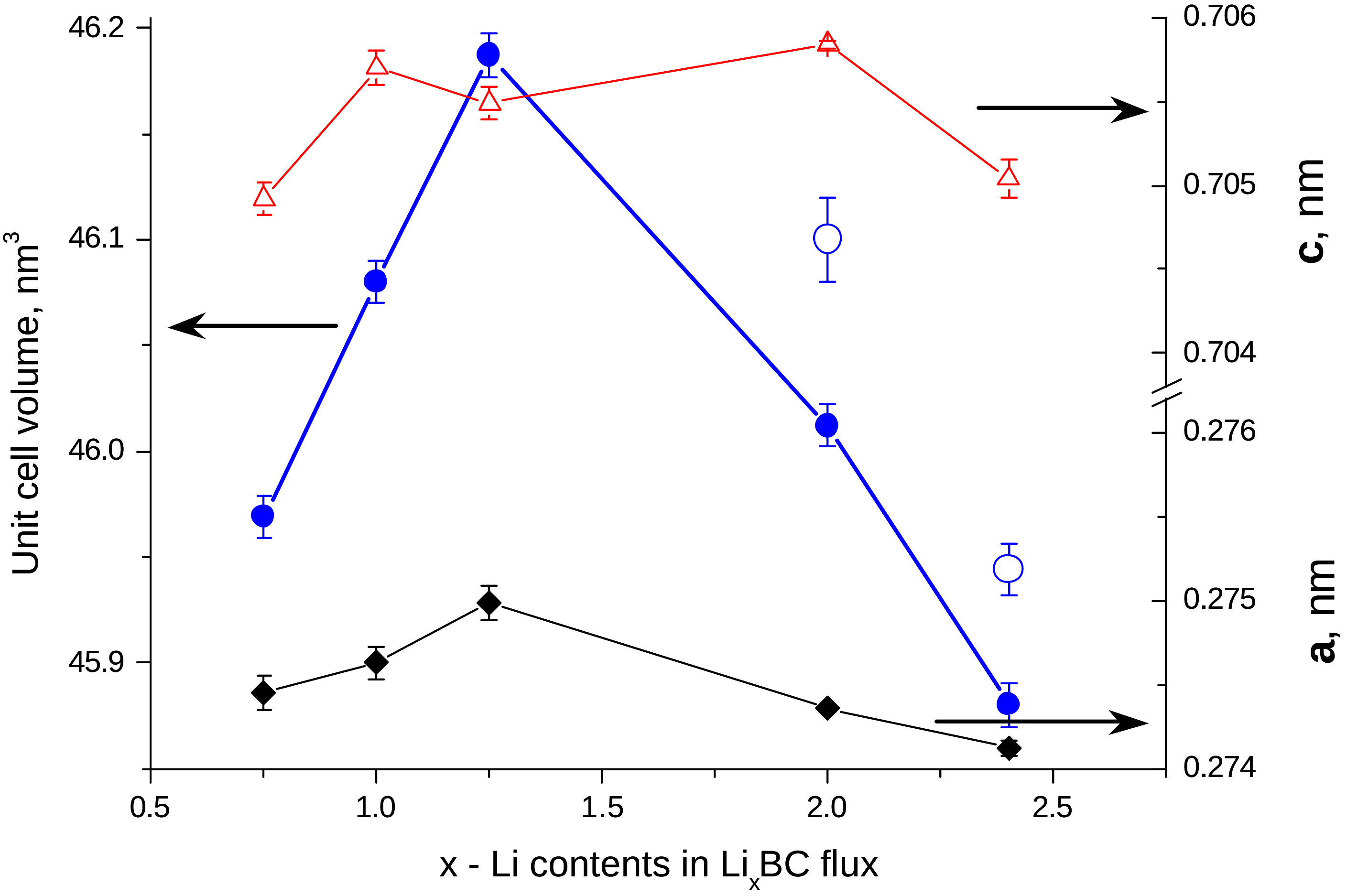

Fig. 3. Lattice parameters a, c and unit cell volume of the LiBC compound as function of the Li-fraction *x* of the $Li_xBC$ melt flux. Open dots – after annealing 900°C for 20 h.

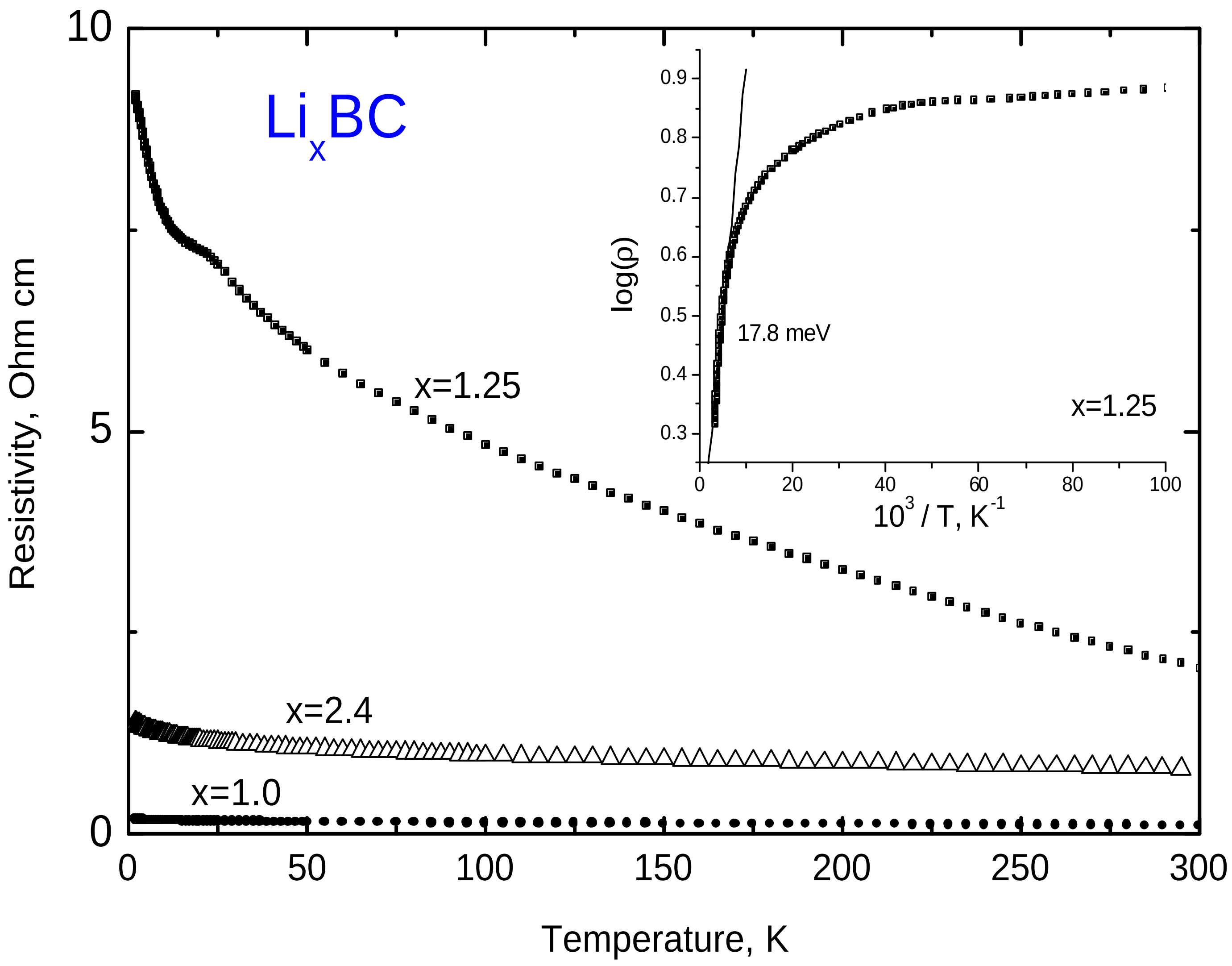

Fig. 4. Temperature dependence of the electrical resistivity ρ of LiBC samples prepared from various flux compositions $Li_xBC$, $x$ = 1.0, 1.25, and = 2.4. Inset: Log ρ vs. 1/T of the single-phase sample ($x$ = 1.25).

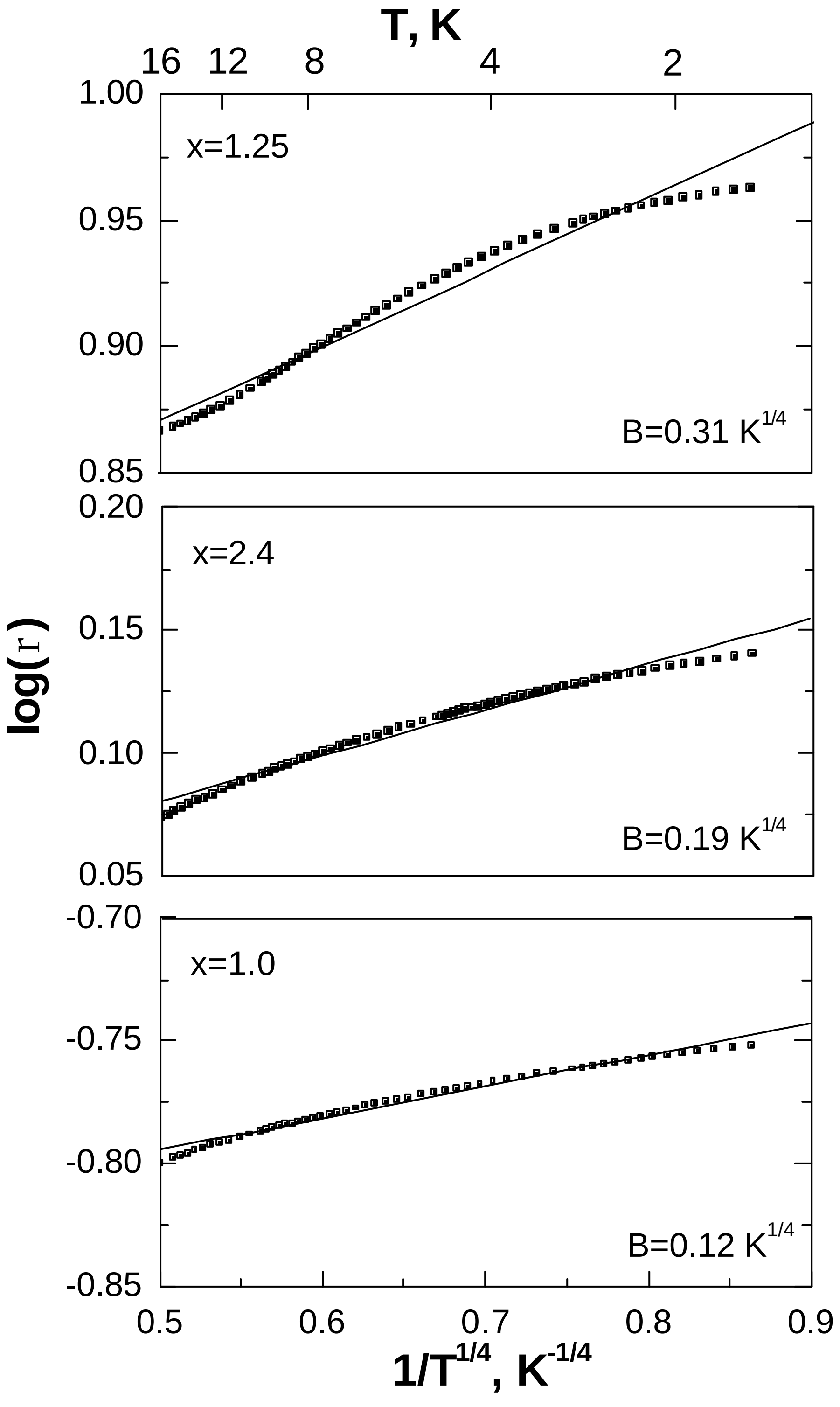

Fig. 5. Log ρ vs. $T^{1/4}$ plot of the electrical resistivity ρ at T ≤ 16 K for LiBC samples prepared from various flux compositions $Li_xBC$, *x* = 1.0, 1.25, and = 2.4. Full line: linear fit.

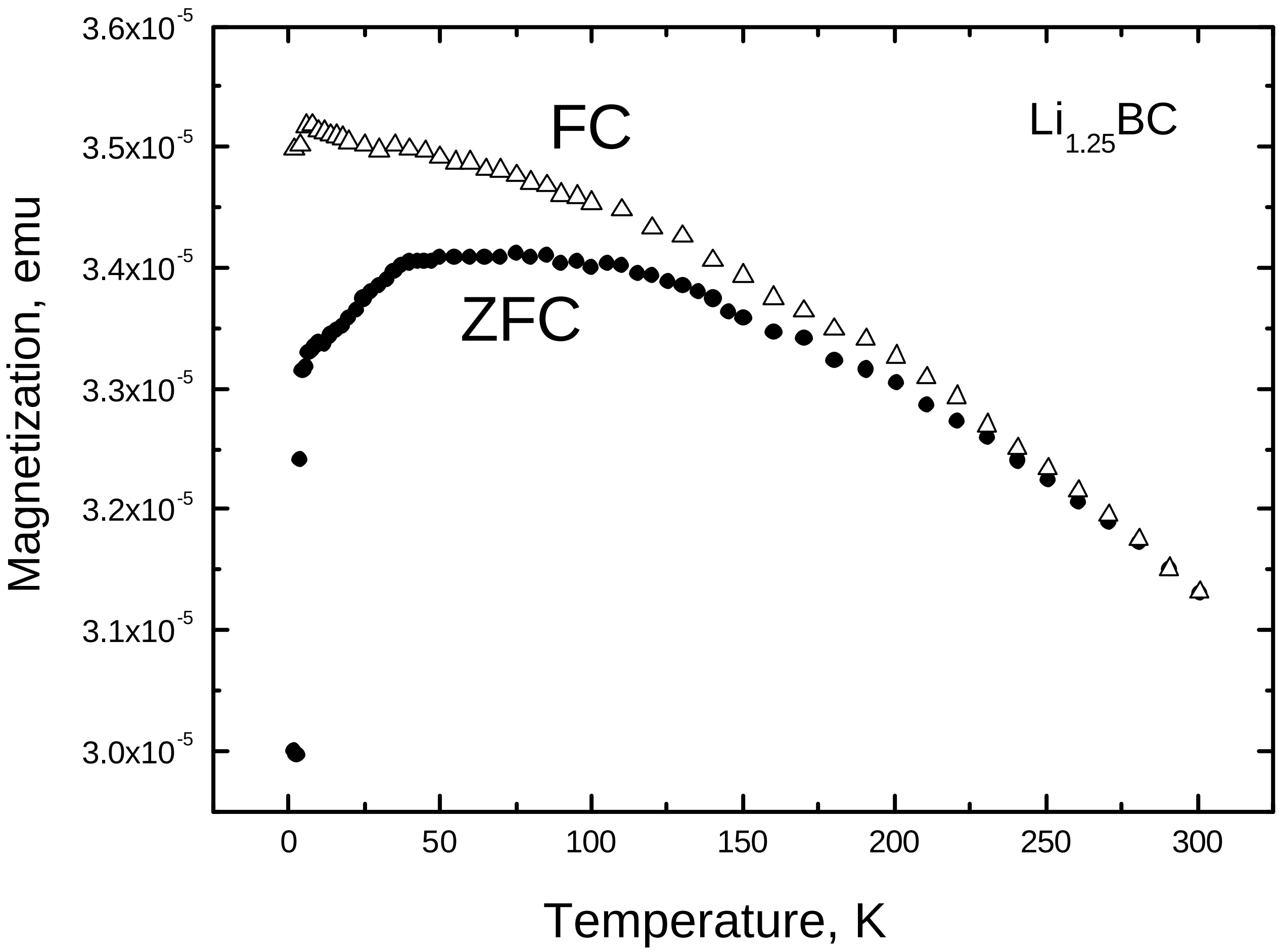

Fig. 6. Temperature dependence of magnetization for LiBC samples prepared from a $Li_{1.25}BC$ flux for zero-field cooled (ZFC) and field cooled (2 mT, FC) conditions.